\documentclass[english]{article}
\usepackage[T1]{fontenc}
\usepackage[latin9]{inputenc}
\usepackage{float}
\usepackage{amsthm}
\usepackage{amsmath}
\usepackage{graphicx}
\usepackage{amssymb}

\makeatletter
\newcommand{\lyxaddress}[1]{
\par {\raggedright #1
\vspace{1.4em}
\noindent\par}
}

\makeatother

\usepackage{babel}

\begin{document}

\title{Stochastic dynamics of an electron in a Penning trap: phase flips
correlated with amplitude collapses and revivals}

\author{S. Brouard and J. Plata}

\maketitle

\lyxaddress{\begin{center}
Departamento de F\'{\i}sica Fundamental II, Universidad de La Laguna,\\
 La Laguna E38204, Tenerife, Spain.
\par\end{center}}
\begin{abstract}
We study the effect of noise on the axial mode of an electron in a
Penning trap under parametric-resonance conditions. Our approach,
based on the application of averaging techniques to the description
of the dynamics, provides an understanding of the random phase flips
detected in recent experiments. The observed correlation between the
phase jumps and the amplitude collapses is explained. Moreover, we
discuss the actual relevance of noise color to the identified phase-switching
mechanism. Our approach is then generalized to analyze the persistence
of the stochastic phase flips in the dynamics of a cloud of $N$ electrons.
In particular, we characterize the detected scaling of the phase-jump
rate with the number of electrons. 
\end{abstract}

\section{Introduction}

The research on electron traps has opened the way to significant advances
in fields ranging from Atomic Physics to Metrology \cite{key-1}.
For instance, the application of the trapping techniques has been
crucial for achievements like the generation of antimatter atoms \cite{key-2}
or the realization of precision tests on fundamental constants \cite{key-3}
Moreover, trapped electrons provide a controllable testing ground
for a variety of physical behaviors, predicted or experimentally identified
in other areas. Actually, the possibility of controlling the trapping
setup, in particular, of varying its components, can allow the systematic
characterization of different effects via their realization under
well-defined conditions and in regimes unexplored in other contexts.
In this way, problems like the emergence of nontrivial effects of
noise \cite{key-4}, the preparation of Fock states \cite{key-5},
the appearance of squeezing in quantum-dissipation processes in nonlinear
oscillators \cite{key-6,key-7,key-8}, or the implementation of proposals
for quantum-information algorithms \cite{key-9,key-10,key-11,key-12}
have been analyzed with different variations of the basic trapping
setup. Here, we focus on a novel effect detected in recent experiments
on electrons in a Penning trap \cite{key-4}. Namely, under parametric-resonance
conditions, the axial mode of a one-electron system was observed to
present random amplitude collapses correlated with phase flips. That
behavior was traced to noise rooted in different elements of the practical
arrangement. Indeed, by adding fluctuations in a controlled way, the
dependence of the phase-jump rate on the noise strength was characterized.
Remarkably, for increasing noise intensities, the correlation between
the amplitude collapses and the phase jumps was found to disappear.
The study of the persistence of those effects in the dynamics of a
cloud of $N$ electrons revealed a nontrivial behavior, in particular,
the attenuation and eventual disappearance of the stochastic phase
switching as the number of electrons was increased. Despite the advances
in the characterization of the observed dynamics \cite{key-4,key-13},
a satisfactory explanation of the underlaying physical mechanisms
is still needed, as stressed in Ref. {[}4{]}. For example, the actual
relevance of colored noise to the emergence of some of the detected
effects is an open question. Here, we present a description of the
dynamics based on averaging techniques applicable to stochastic systems.
From our approach, the origin of the experimental findings is uncovered
and the elements of the system which are essential for the appearance
of the observed behavior are identified. Furthermore, our semi-analytical
characterization of the dynamics provides us with some clues to controlling
the system response to the fluctuations. The direct implications of
the study to the advances in the techniques of confinement and stabilization
are evident. Additionally, because of the fundamental character of
the physics involved, the analysis can be relevant to different contexts,
where conditions similar to those realized in the trapping scenario
can be implemented \cite{key-14,key-15}.

The outline of the paper is as follows. In Sec. II, we present our
model for the stochastic dynamics of the axial mode of an electron
in a Penning trap. Through the application of the averaging methods
of Bogoliubov, Krylov, and Stratonovich \cite{key-16,key-17}, we
derive an effective description in terms of a system of stochastic
differential equations for the amplitude and the phase. In Sec. III,
the validity of our approach is confirmed through the simulation of
the main experimental findings. Moreover, an analysis of the physical
mechanisms responsible for the observed behavior is presented. Sec.
IV contains the study of the persistence of the stochastic phase switching
in the dynamics of a cloud of $N$ electrons. Finally, some general
conclusions are summarized in Sec. V.

\section{The model system}

An electron in a Penning trap is usually described in terms of three
coupled modes (magnetron, axial, and cyclotron) with widely different
time scales \cite{key-1}. Here, we concentrate on the axial coordinate
$z$.  In general, because of the intermode coupling, the axial dynamics
can be quite complex: a variety of behaviors can emerge depending
on the considered regime of experimental parameters. However, under
standard conditions, different approximations can be made and the
description simplifies considerably. Namely, the (slow) magnetron
motion can be adiabatically treated and its influence on the dynamics
can be reduced by sideband cooling \cite{key-4}. Additionally, by
damping the (fast) cyclotron mode to its fundamental state, its effect
on the axial coordinate is minimized \cite{key-4}. In typical realizations,
$z$ is coupled to a measuring external circuit, which introduces
resistive damping and noise in the mode; moreover, the trap potential
that is usually applied is approximately harmonic. Hence, in the basic
scheme, $z$ corresponds to a dissipative harmonic oscillator which
can be described classically. Axial outputs qualitatively different
from that basic realization can be generated by introducing different
driving fields and controlled fluctuations in the practical setup
\cite{key-1,key-18,key-19,key-20}. Apart from directly affecting
the dynamics, those external elements can indirectly enhance the effect
of the nonlinear corrections to the trap potential. Here, we aim at
explaining the experiments of Ref. {[}4{]}. In them, a nontrivial
axial dynamics was uncovered in a variation of the basic setup which
incorporated a driving field at parametric resonance and noise of
controllable intensity. Those experimental conditions are simulated
in our approach. Specifically, we consider that $z$ is parametrically
driven at a frequency $\omega_{d}$ which is nearly twice the characteristic
resonant frequency $\omega_{z}$, i.e., $\omega_{d}=2(\omega_{z}+\epsilon)$
with the restriction $\epsilon\ll\omega_{z}$ for the detuning $\epsilon$.
Moreover, our model incorporates a stochastic force $\eta(t)$ with
the characteristics of the noise present in the practical setup. Residual
nonlinear terms of the confining potential, which are known to account
for the stabilization of the noiseless version of the system in the
parametric-resonance regime \cite{key-18,key-19}, are also included
in the model. Accordingly, we consider that the  axial coordinate,
(normalized to a typical trap length, and, therefore, dimensionless),
is described by the equation

\begin{equation}
\ddot{z}+\gamma_{z}\dot{z}+\omega_{z}^{2}\left[1+h\cos\omega_{d}t\right]z+\lambda_{4}\omega_{z}^{2}z^{3}+\lambda_{6}\omega_{z}^{2}z^{5}=\eta(t),\end{equation}
where $\gamma_{z}$ is the friction coefficient, $h$ characterizes
the amplitude of the driving force, and, $\lambda_{4}$ and $\lambda_{6}$
are coefficients which determine the magnitude of the nonlinear terms
of the confining potential. By now, we deal with a one-electron system;
later on, we will tackle the dynamics of a cloud of $N$ electrons.

Crucial to the applicability of our model to the considered experiments
is the appropriate modeling of the stochastic force $\eta(t)$. Two
types of fluctuations are relevant to the experimental scheme. First,
the system presents {}``internal'' noise rooted in different elements
of the practical setup. It has been argued that those fluctuations
are well modeled by broadband noise and centered narrow-band fluctuations.
Second, white noise, more intense than the internal fluctuations,
was injected in the experimental realization of Ref. {[}4{]}. Indeed,
it was the addition of this {}``external'' noise that allowed studying
the dependence of the switching mechanism on the noise strength. Here,
in order to account for both, the broadband internal fluctuations
and the added white noise, we consider that $\eta(t)$ has general
Gaussian wideband characteristics \cite{key-21}. Specifically, we
assume that the correlation function $k_{\eta}(t^{\prime}-t)\equiv\left\langle \eta(t)\eta(t^{\prime})\right\rangle -\left\langle \eta(t)\right\rangle ^{2}$
has a generic functional form and that the correlation time is much
shorter than any other relevant time scale in the system evolution.
The intensity coefficient $D=\frac{1}{2}\int_{-\infty}^{\infty}k_{\eta}(\tau)d\tau$
will be used to characterize the noise strength \cite{key-17}. (The
white-noise limit, defined by $k_{\eta}(t^{\prime}-t)=2D\delta(t-t^{\prime})$,
is included in our analysis.) Additionally, a zero mean value, $\left\langle \eta(t)\right\rangle =0$,
is assumed. (Notice that a nonzero $\left\langle \eta(t)\right\rangle $
can be simply incorporated into the model as an effective deterministic
contribution.) A more elaborate noisy input should be added to tackle
the effect of residual colored noise. However, that generalization
of our approach is not necessary for the objectives of the present
paper: it will be shown that the detected behavior can be simply traced
to broadband-noise characteristics. 

Our approach to deal with Eq. (1) is based on the averaging methods
developed by Krylov and Bogoliubov for the analysis of deterministic
nonlinear oscillations as they were generalized by Stratonovich to
the study of stochastic processes \cite{key-16,key-17}. Those averaging
techniques can be applied to generic wideband fluctuations with sufficiently
short correlation time. In this approach, the amplitude $A$ and the
phase $\Psi$ of the oscillations are defined through the equations
$z=A\cos\left[(\omega_{z}+\epsilon)t+\Psi\right]$ and $\dot{z}=-(\omega_{z}+\epsilon)A\sin\left[(\omega_{z}+\epsilon)t+\Psi\right]$.
With these changes, Eq. (1) is reduced to a system of two first-order
equations in \textit{standard form} \cite{key-17}, i.e., with the
structure of a harmonic oscillator perturbed by deterministic and
stochastic terms. For $\omega_{z}\gg\gamma_{z}$, $\epsilon$, the
average of the deterministic perturbative elements over the period
$\tau_{ef}=2\pi/(\omega_{z}+\epsilon)=4\pi/\omega_{d}$ is readily
carried out. Moreover, for a noise correlation time much smaller than
the relaxation times of the amplitude and the phase, the coarse graining
of the stochastic terms over $\tau_{ef}$ can be applied following
the procedure presented in Ref. {[}17{]}. Accordingly, we obtain that,
to first-order, the averaged equations are \cite{key-17,key-20,key-22}

\begin{equation}
\dot{A}=-\frac{\gamma_{z}}{2}[1-\frac{h}{h_{T}}\sin2\Psi]A+\frac{D_{eff}}{A}+\xi_{1}(t),\end{equation}

\noindent \begin{equation}
\dot{\Psi}=-\epsilon+\frac{3}{8}\lambda_{4}\omega_{z}A^{2}+\frac{5}{16}\lambda_{6}\omega_{z}A^{4}+\frac{1}{4}\omega_{z}h\cos2\Psi+\frac{\xi_{2}(t)}{A},\end{equation}
where we have introduced $h_{T}\equiv2\gamma_{z}/\omega_{z}$. (The
meaning of $h_{T}$ as a threshold amplitude of the driving field
will be evident shortly.) Additionally, $\xi_{1}(t)$ and $\xi_{2}(t)$
are effective Gaussian white-noise terms defined by $\left\langle \xi_{i}(t)\right\rangle =0$,
and $\left\langle \xi_{i}(t)\xi_{j}(t^{\prime})\right\rangle =2D_{eff}\delta_{i,j}\delta(t-t^{\prime})$,
$i,j=1,2$, with $D_{eff}=\kappa_{\eta}(\omega_{z}+\epsilon)/[4(\omega_{z}+\epsilon)^{2}]$.
(Note that $D_{eff}$, which determines the strength of the (uncorrelated)
effective noise terms, is obtained from the power spectral density
$\kappa_{\eta}(\omega)\equiv\int_{-\infty}^{\infty}e^{i\omega\tau}k_{\eta}(\tau)d\tau$
of the original noise $\eta(t)$ at the frequency $\omega_{z}+\epsilon$.
Here, we must remark that, from the broadband characteristics assumed
for $\eta(t)$, a smooth form of $\kappa_{\eta}(\omega)$ can be inferred.
Indeed, a completely flat spectrum occurs in the white-noise limit.)
Whereas the noise term in Eq. (2), $\xi_{1}(t)$, is additive, the
fluctuations enter Eq. (3) through the term $\xi_{2}(t)/A$, and,
therefore, have multiplicative character. Moreover, it is important
to take into account the presence of the noise-induced {}``deterministic''
term $D_{eff}/A$ in Eq. (2): its appearance will be shown to account
for the partial character of the amplitude collapses detected in the
experiments. It is worth emphasizing that our use of averaged equations
is specially appropriate for the considered experimental setup, where,
because of the specific characteristics of the detection scheme, the
registered data do actually correspond to averaged magnitudes. 

In order to trace the response of the system to noise, we must clearly
define the deterministic scenario into which the fluctuations enter.
The noiseless dynamics of the system, described by Eq. (1) without
the random term $\eta(t)$, and, consequently, by Eqs. (2) and (3)
with $D_{eff}=0$, has been intensively studied \cite{key-18,key-19}.
From the averaged equations, it is straightforwardly shown that parametric
amplification, i.e., exponential growth of the amplitude, takes place
for a driving amplitude $h$ larger than the threshold value $h_{T}$
and for a detuning $\epsilon$ within the excitation range, namely,
for $\epsilon_{-}<\epsilon<\epsilon_{+}$ ($\epsilon_{\pm}=\pm\frac{\omega_{z}}{4}\sqrt{h^{2}-h_{T}^{2}}$.)
The experimental conditions on which we focus correspond to this parametric-amplification
regime. In the absence of nonlinear terms in the trap potential, the
amplitude would grow monotonously. However, the nonlinear corrections,
characterized by the coefficients $\lambda_{4}$ and $\lambda_{6}$,
which must be included in the description to simulate the actual potential
applied in the practical setup, do arrest the amplitude growth, allowing
the stabilization of the motion. The system presents two stationary
states with the same amplitude and with phase values differing in
$\pi$ radians. Specifically, the stationary amplitude $A_{SS}$ is
obtained from the equation $\epsilon_{+}-\epsilon+\frac{3}{8}\lambda_{4}\omega_{z}A_{SS}^{2}+\frac{5}{16}\lambda_{6}\omega_{z}A_{SS}^{4}=0$,
and the two $\pi$-differing values of the equilibrium phase $\Psi_{SS}$
are given by $\Psi_{SS}=\frac{1}{2}\arcsin(\frac{h_{T}}{h})$. {[}Important
for the discussion of some of the noisy features is to notice that,
for the values of the nonlinear parameters applied in the experiments,
($\lambda_{4}=0$ and $\lambda_{6}<0$), $A_{SS}$ increases with
$\epsilon_{+}-\epsilon$.{]} The stationary states correspond to attractors
in the phase space. Depending on the initial conditions, the system
eventually reaches one or other attractor. The objective of the next
section is the explanation of the effects of noise on this deterministic
scenario.

\section{Stochastic dynamics of a one-electron system}

In the analysis of the noisy dynamics, we proceed by showing first
that Eqs. (2) and (3) provide a satisfactory description of the behavior
observed in the experiments. Then, once its validity has been confirmed,
our approach will be applied to uncover the physics underlying the
detected features.

\subsection{The response to noise}

In Ref. {[}4{]}, the presence of noise was shown to significantly
alter the deterministic picture. The system was not longer stabilized
in one of the attractors. Instead, it was observed to display amplitude
collapses and revivals correlated with abrupt changes in the phase.
Those experimental findings are reproduced by our approach. Figs.
1a and 1b respectively depict results for $A$ and $\Psi$ as obtained
from Eqs. (2) and (3). There, the correlation between the phase flips
and the collapses and revivals of the amplitude is evident. Actually,
in agreement with the experimental results, it is found that the amplitude
collapses are almost always followed by phase flips. In other words,
the system rarely stays in the same basin of attraction once a collapse
in $A$ has occurred. One should notice that, in the inter-jump intervals,
$\Psi$ is strongly localized around its equilibrium values; in contrast,
a significant dispersion in $A$ is observed. Our study reproduces
the detected partial character of the collapses. Already noticeable
in Fig. 1a, this feature is particularly evident in Fig. 2, where
we depict a typical noisy trajectory, which includes different flips
between the two basins. There, it is apparent that the unstable point
defined by $A=0$ is never reached. As stressed in Ref. {[}4{]}, these
features cannot be understood with a simple activation-process model. 

\includegraphics[angle=270,scale=0.3]{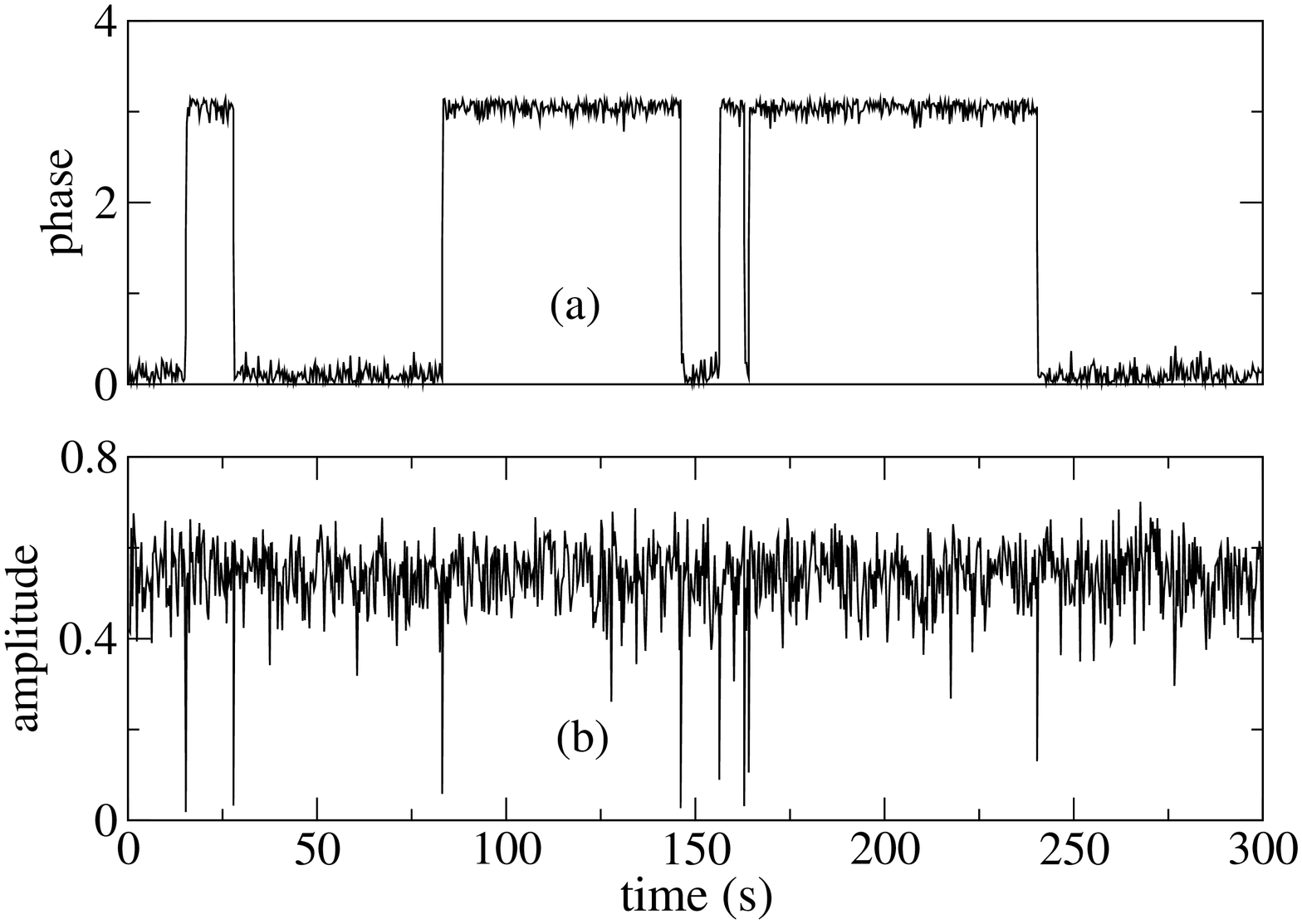}

\begin{figure}[H]
\caption{Time evolution of the phase (a) and of the amplitude (b) as obtained
from Eqs. (2) and (3). {[}$\omega_{z}/2\pi=61.6MHz$, $\lambda_{4}=0$,
$\lambda_{6}=-0.27$, $\gamma_{z}=(10ms)^{-1}$, $\epsilon_{+}/2\pi=100Hz$,
$\epsilon/2\pi=50Hz$, $D=10^{-2}$ (arbitrary units).{]} (The same
set of parameters is used throughout the paper.) }

\end{figure}

\includegraphics[angle=270,scale=0.3]{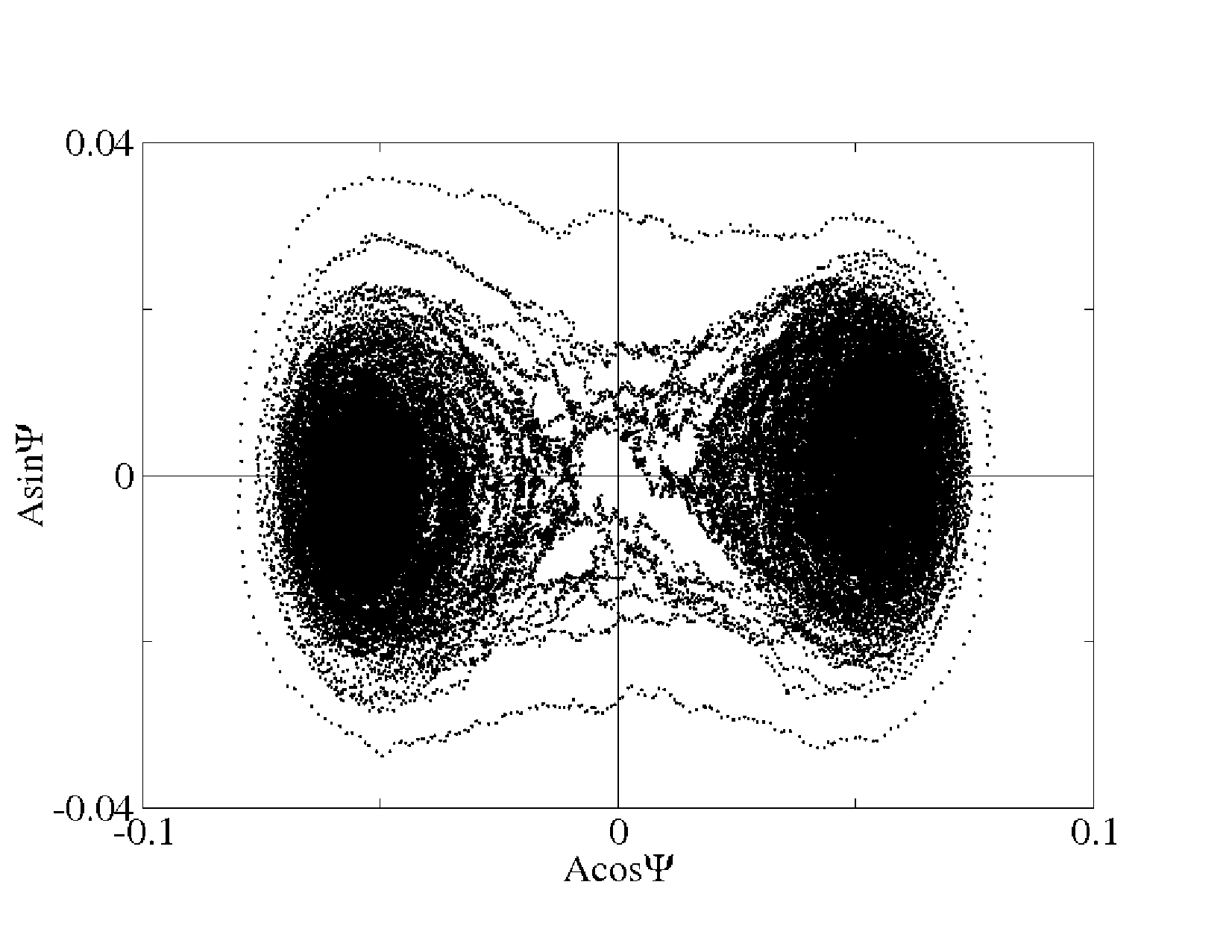}

\begin{figure}[H]
\caption{Phase space diagram for a particular noisy trajectory. }

\end{figure}

In the analysis of the experimental results, a histogram for the residence
time, (i.e., the time interval between phase flips), served to obtain
the average jump rate $\Gamma$. Actually, $\Gamma$ was found to
be well approximated by an exponential function of the noise strength,
namely, $\Gamma\thicksim\exp(-E/D)$, where $E$ denotes the effective
activation energy. That characterization is reproduced by our approach.
In Fig. 3, we represent a histogram for the residence time as obtained
from Eqs. (2) and (3). Moreover, in Fig. 4, we plot the jump rate
as a function of the noise strength. There, the validity of the exponential
fit is patent. (In our calculations, we have directly worked with
$D_{eff}$ as noise strength: the used arbitrary units include the
ratio between the actual strength of the original noise $D$ and $D_{eff}$.)
Apart from the dependence on the fluctuations, the jump rate incorporates,
through the effective activation energy, the influence on the process
of elements like the frequency and strength of the driving field,
the characteristics of the trapping potential, or the damping coefficient.
The description of the role played by those deterministic components
of the system in the noisy dynamics is crucial for understanding the
process of phase switching. In the experiments, the dependence of
$E$ on those system parameters was traced via the systematic variation
of the practical conditions. In particular, an approximately exponential
dependence of the jump rate $\Gamma$ on the detuning, expressed as
$\epsilon_{+}-\epsilon$, was reported. That behavior is also simulated
with our approach, as shown in Fig. 5.

\includegraphics[angle=270,scale=0.3]{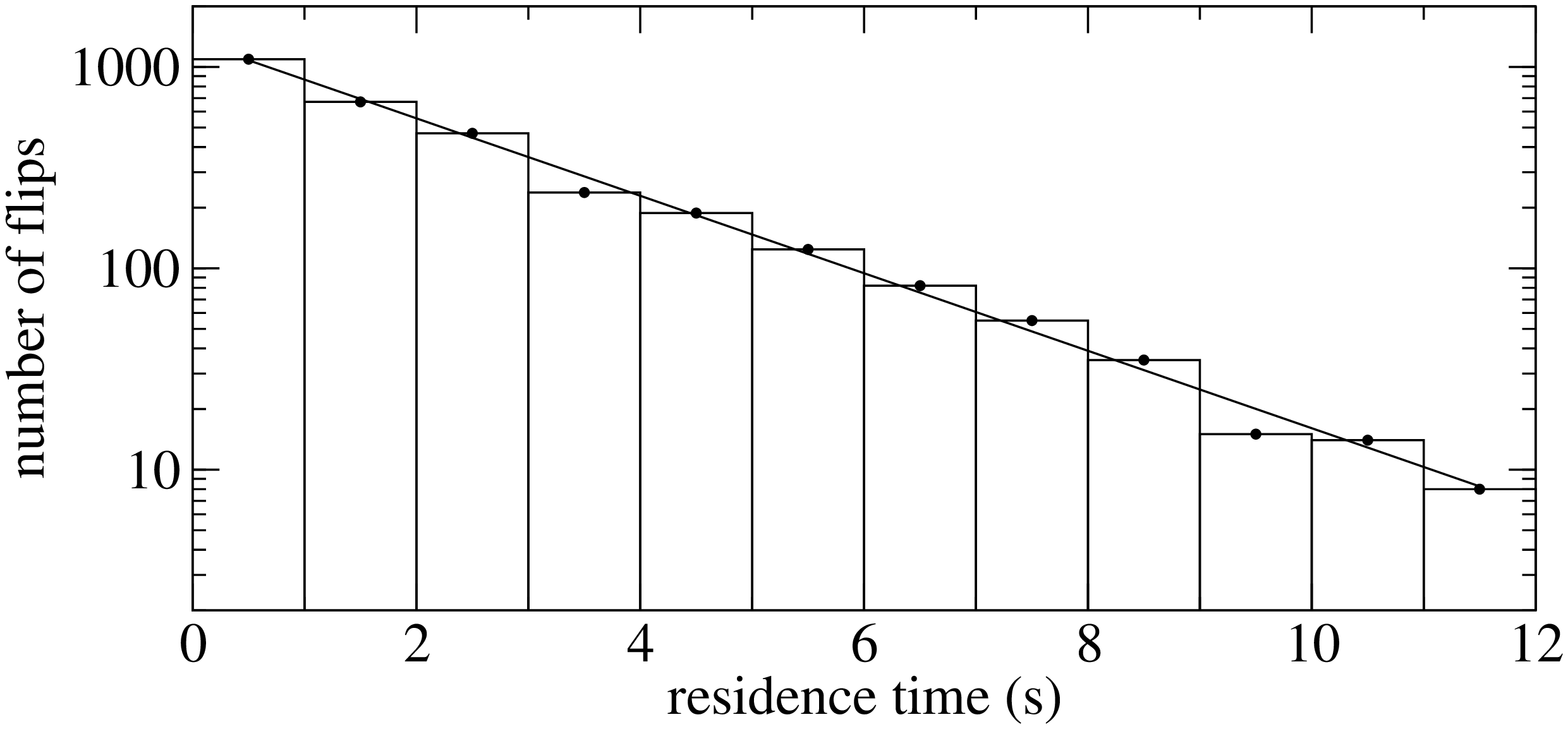}

\begin{figure}[H]
\caption{Histogram for the time interval between phase flips.}

\end{figure}

\includegraphics[angle=270,scale=0.3]{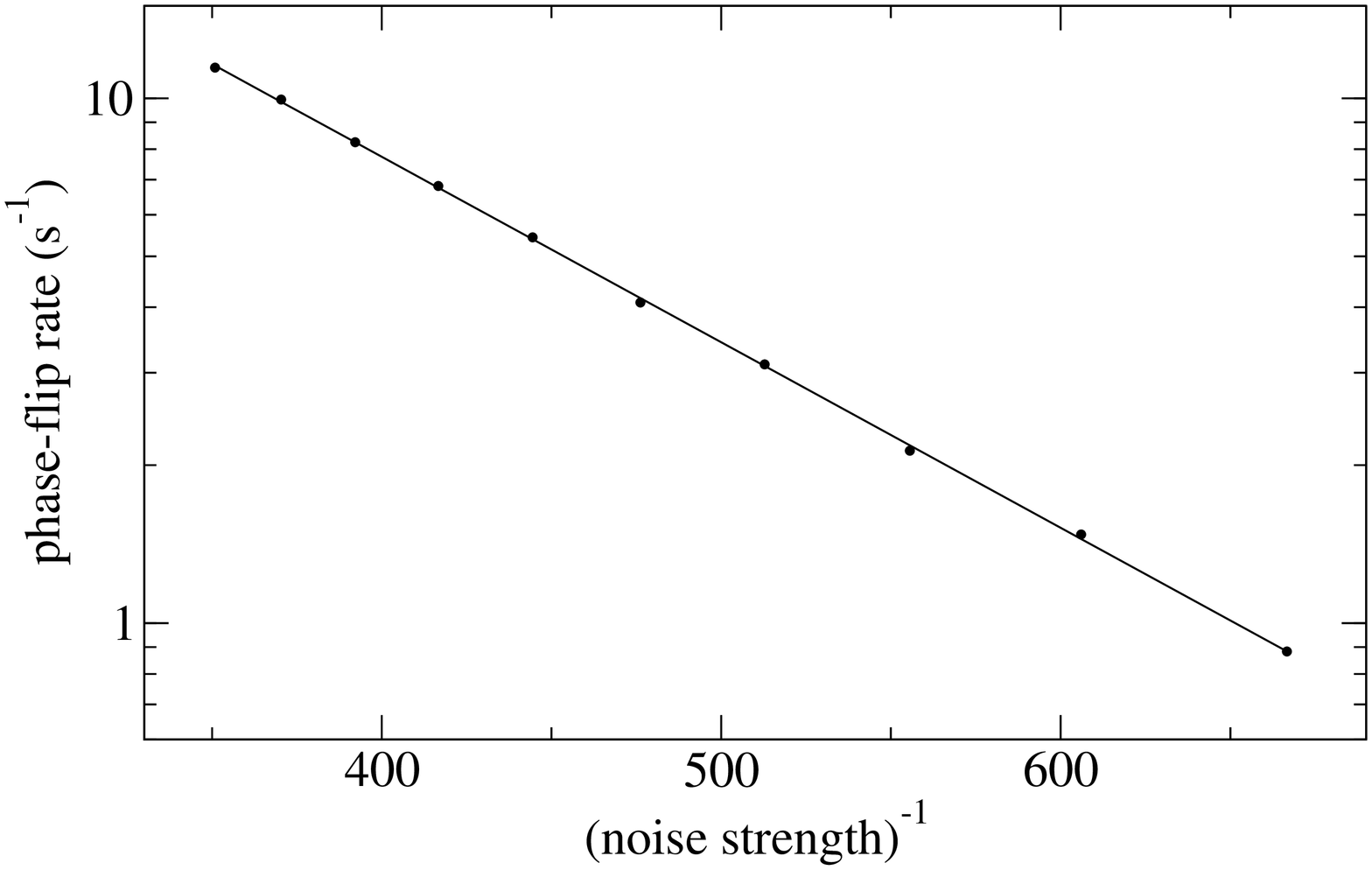}

\begin{figure}[H]
\caption{Phase-jump rate $\Gamma$($s^{-1}$) versus the inverse noise-strength
$D^{-1}$ (arbitrary units).}

\end{figure}

\includegraphics[angle=270,scale=0.3]{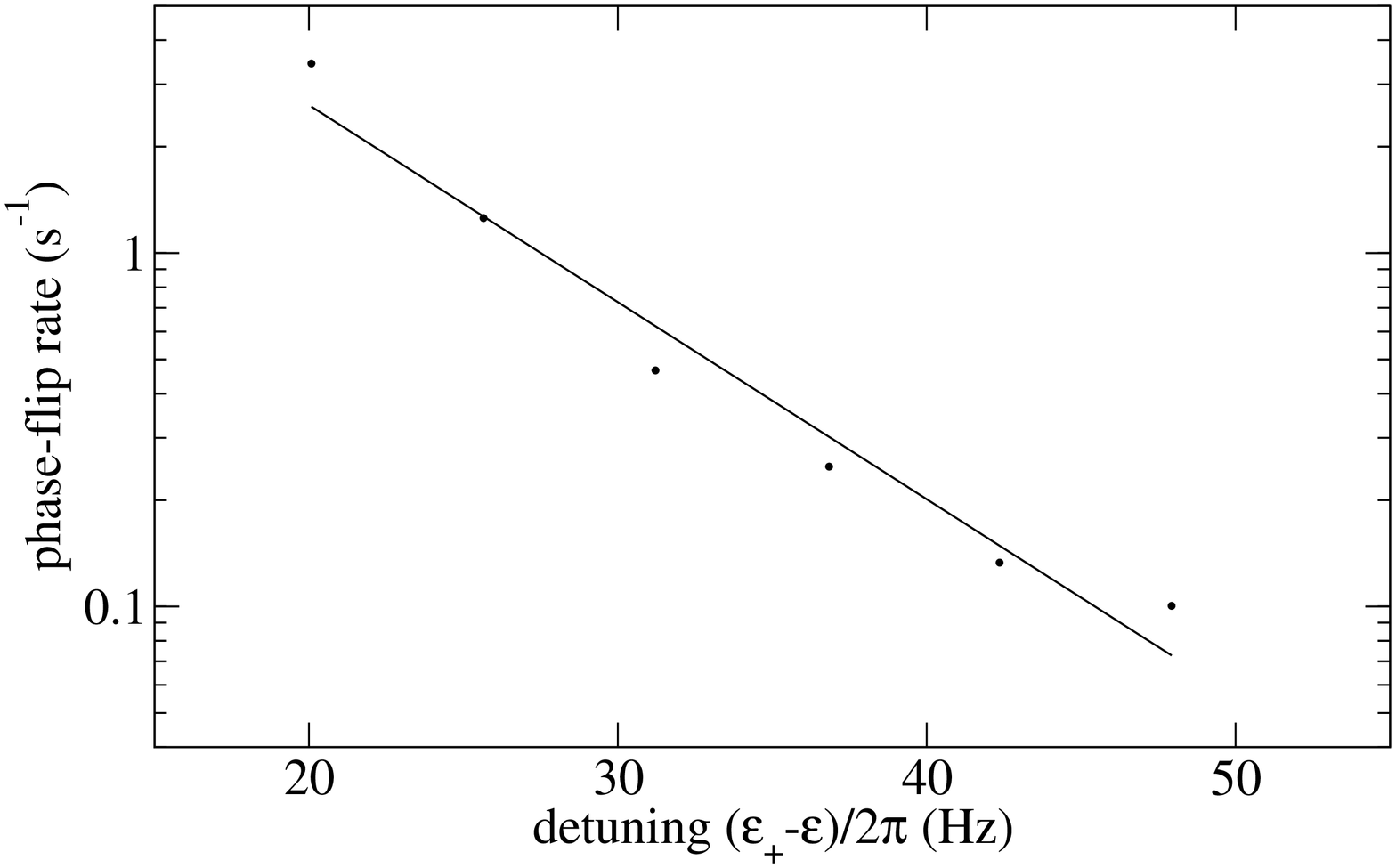}

\begin{figure}[H]
\caption{Phase-jump rate $\Gamma$($s^{-1}$) as a function of the detuning
expressed as $(\epsilon_{+}-\epsilon)/2\pi$ ($Hz$). ($\epsilon_{+}/2\pi=100Hz$). }

\end{figure}

In Ref. {[}4{]}, no results are presented for the dependence of the
activation energy on the friction coefficient. However, as this aspect
of the system behavior will be an important element of our discussion
of the phase switching mechanism, it is pertinent to describe it here
using our approach. Indeed, since early theoretical studies were set
up from a Hamiltonian approximation to the dynamics, it is worthwhile
to inquire into the actual significance of the dissipative character
of the system. Let us first recall some aspects of the purely deterministic
(dissipative) dynamics which are relevant to the resulting stochastic
scenario. Namely, from the found threshold amplitude, given by $h_{T}=2\gamma_{z}/\omega_{z}$,
it is evident that the generation of oscillations is inhibited as
the friction coefficient $\gamma_{z}$ increases. Additionally, we
must take into account that, as found in Ref. {[}11{]}, the relaxation
of the system from any initial conditions to the equilibrium states
becomes faster for larger $\gamma_{z}$. Therefore, we can conjecture
that, as $\gamma_{z}$ grows, simply because of the deterministic
inhibition of the oscillations and of the enhanced stability of the
system, the role of noise in activating the phase jumps must be hindered.
This conjecture is confirmed by our results for the dependence of
$E$ on $\gamma_{z}$. We have found that there is an approximately
exponential decrease of the jump rate $\Gamma$ with $\gamma_{z}$,
as shown in Fig. 6. Consequently, from the expression $\Gamma\thicksim\exp(-E/D)$,
a nearly linear increase of $E$ with $\gamma_{z}$ is derived. Specifically,
we can write $E\approx C_{1}+C_{2}\gamma_{z}$, where the coefficients
$C_{1}$ and $C_{2}$ incorporate the dependence of $E$ on other
parameters of the system. Some implications of these results will
be considered in the forthcoming discussion. By now, we anticipate
that the analysis of the persistence in the $N$-electron system of
the found dependence of $E$ on $\gamma_{z}$ will be central to our
understanding of the detected scaling of the phase jump rate with
$N$.

\includegraphics[angle=270,scale=0.3]{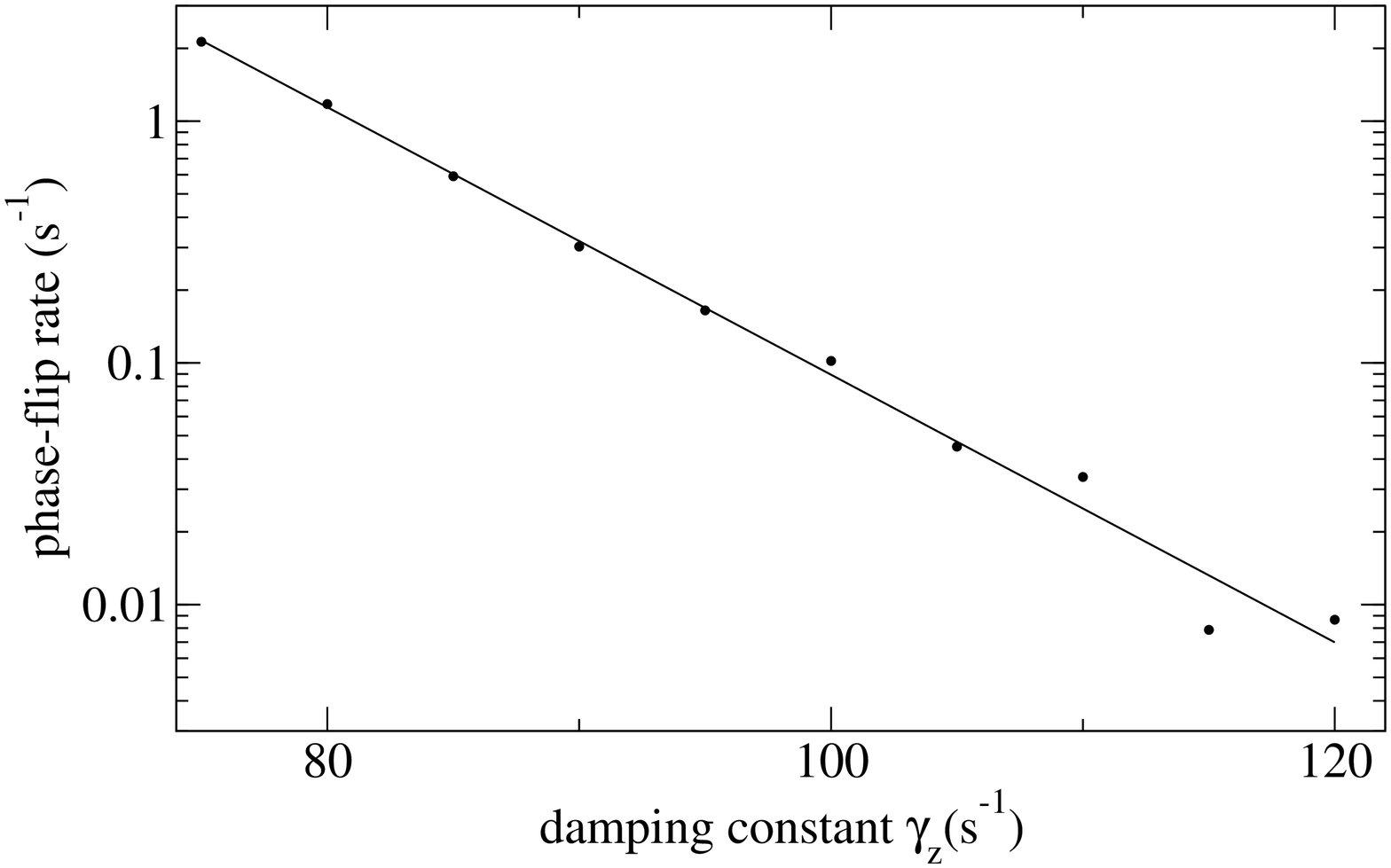}

\begin{figure}[H]
\caption{Phase-jump rate $\Gamma$($s^{-1}$) versus the friction coefficient
$\gamma_{z}$ ($s^{-1}$).}

\end{figure}

\subsection{The mechanism of stochastic phase switching}

In our discussion of the physics that underlies the observed features,
we proceed gradually: we start with a simplified picture of the dynamics,
which will be improved by successively incorporating the different
elements of the complete system. Our scheme is summarized in the following
steps.

(i) A zero-order approximation to the random response detected in
the experiments is provided by the artificial decoupling of Eqs. (2)
and (3). Indeed, some clues to the origin of prominent features of
the complete system are given by the analysis of the {}``independent''
behaviors of $A$ and $\Psi$ that respectively follow from fixing
the phase in Eq. (2) and the amplitude in Eq. (3). 

For a constant value of $\Psi$, Eq. (2) describes fluctuations of
$A$ around its equilibrium position. Interestingly, because of the
term $D_{eff}/A$, the equilibrium amplitude is larger than its counterpart
in the absence of noise. Noise-induced excursions to the region of
small $A$ can be predicted. Due to the effective {}``deterministic''
term $D_{eff}/A$, the value $A=0$ is not reached, i.e., a complete
collapse never takes place. As in the description of the deterministic
system at parametric resonance, to account in our approach for the
limited growth of $A$ observed in practice, the coupling to the phase
equation, which contains the nonlinear terms of the confining potential,
is necessary.

Conversely, for a fixed $A$, Eq. (3) describes a process of phase
diffusion in a tilted periodic potential \cite{key-17}. The bias,
given by $-\epsilon+\frac{3}{8}\lambda_{4}\omega_{z}A^{2}+\frac{5}{16}\lambda_{6}\omega_{z}A^{4}$,
is determined by the detuning and by the artificially fixed amplitude.
The potential presents two minima separated by $\pi$ radians.When
the bias is smaller than the height of the periodic potential, $\Psi$
evolves only because of the fluctuations. Actually, in the regime
considered in the experiments, it is noise that leads to phase jumps
between the minima. Since the magnitude of the (multiplicative) random
term $\xi_{2}(t)/A$ increases as $A$ diminishes, the jumps become
more frequent for smaller $A$. 

(ii) A comparative analysis of the structure of the two averaged equations
uncovers the qualitatively different effects of noise on the two variables.
Since the amplitude has no strong confining potential, it is continuously
forced out of equilibrium by the stochastic force. Indeed, a significant
dispersion in $A$ is apparent in Fig. 1a. In contrast, as shown in
Fig. 1b, the periodic potential leads to a remarkable concentration
of $\Psi$ around its equilibrium values. This phase locking is interrupted
by noise-induced flips. Attention must also be paid to some characteristics
of the coupling between Eqs. (2) and (3). As previously discussed,
the presence of $A$ in Eq. (3) is crucial for the evolution of $\Psi$.
In particular, the magnitude of $A$ determines the frequency of the
phase flips via the random term $\xi_{2}(t)/A$. In contrast, a much
weaker effect of the phase flips on the evolution of the amplitude
is apparent: given that the phase enters Eq. (2) through $\sin2\Psi$,
the $\pi$ jumps in the phase hardly alter the dynamics of the amplitude. 

(iii) By combining the ideas contained in the above points, the observed
correlation between amplitude collapses and phase jumps can be explained.
Noise can induce an appreciable reduction in the amplitude, which
leads to a significant increase of the random term $\xi_{2}(t)/A$
in the equation for the phase evolution, and, in turn, to a stochastic
phase jump. Because of the {}``deterministic'' term $D_{eff}/A$,
the unstable point with $A=0$ and undefined $\Psi$ is avoided in
the switching. In fact, as $A$ never reaches a zero value, the phase
is always well-defined. Additionally, the fast regrowth of the amplitude
after each (partial) collapse is rooted in the term $D_{eff}/A$ and
in the friction-induced stability of the underlying deterministic
attractors. In the experiments, the correlation between the amplitude
collapses and the phase jumps was observed to decay for increasing
noise intensities. This feature can be understood taking into account
that, as the noise strength increases, the random term $\xi_{2}(t)/A$
can be strong enough to lead to phase flips even without an appreciable
reduction in $A$. Furthermore, the inhibition of the jumps for increasing
$\gamma_{z}$ is linked to the higher stability of the (deterministic)
stationary amplitude $A_{SS}$ and to the consequent less probable
exploration by the system of the small-amplitude region. Then, we
can understand that, as shown in Fig. 6, the phase jumps dwindle as
$\gamma_{z}$ is enhanced, and, correspondingly, that the effective
activation energy increases with $\gamma_{z}$. A similar argument
qualitatively explains the dependence of the flip rate on the detuning,
reflected in Fig. 5. As previously pointed out, $A_{SS}$ grows with
$\epsilon_{+}-\epsilon$; consequently, for increasing $\epsilon_{+}-\epsilon$,
the collapse region is less easily reached, and, in turn, the phase
jumps become less probable. 

From the above discussion, the essential components of the mechanism
responsible for the stochastic phase-switching can be identified.
The periodic potential, rooted in the driving field at parametric
resonance, allows the strong localization of $\Psi$, and, therefore,
the well-defined character of the jumps in phase. Additionally, the
random term $\xi_{2}(t)/A$ accounts for the correlation between amplitude
collapses and phase flips. This stochastic link can be traced to two
fundamental characteristics of the system. First, it is rooted in
the additive character of the input noise $\eta(t)$: with the change
of representation, from $z$ to $A$ and $\Psi$, the fluctuations
become multiplicative and the random connection does appear. Second,
its specific compact form $\xi_{2}(t)/A$, uncovered by the application
of the averaging methods, is a consequence of the broadband-noise
characteristics, which guarantee the applicability of that methodology.
Given the generality of its origin, this random link can be expected
to be relevant to quite generic stochastic oscillators. In fact, it
has been previously characterized in pioneering work on nonlinear
self-excited oscillations in electronic devices \cite{key-17}. Its
intense differential effect on the current scenario results from its
combination with the driving field: as the phase is strongly confined,
its noise-induced evolution, (enhanced for small values of the amplitude),
occurs basically through noticeable jumps between approximate equilibrium
values.  A comment on the role played by the nonlinearity of the system
is also pertinent. It is important to emphasize that the nonlinear
terms of the trap potential, which are necessary for the stabilization
of the amplitude in the parametric-resonance regime, are not essential
components of the phase-switching mechanism. It is the nonlinearity
induced by the parametric driving field, i.e., the phase bi-stability,
that really counts in the process. Also, it is interesting to examine
what can be extracted from our approach about the actual relevance
of colored noise. We recall that the possibility of tracing some of
the observed features to noise-color characteristics was pointed out
in early discussions of the experiments. In this sense, our study
conclusively shows that the experimental features reported in Ref.
{[}4{]} can be induced purely by broadband noise. Even more, we have
found that simple white noise can account for those effects. Finally,
we remark that, as stressed in Ref. {[}4{]}, a simple activation-process
model \cite{key-21} fails to provide an appropriate picture of the
observed dynamics. Although the introduction of an effective activation
energy is useful in the characterization of the phase-jump rate, a
simple activation-process description misses the sequence of combined
effects in the evolution of the amplitude and phase that leads to
the distinctive characteristics of the phase switching.

\section{stochastic dynamics of a cloud of N electrons}

In the above, we have considered a mono-electronic system. Now, we
turn to analyze the dynamics of a cloud of $N$ electrons. We aim
at explaining the nontrivial features of the evolution of the center-of-mass
coordinate $Z$ ($Z=\frac{1}{N}\sum_{i=1}^{N}z_{i}$) uncovered by
the experiments of Ref. {[}4{]}, specifically, the observed slow-down
and eventual disappearance of the random phase jumps for increasing
number of electrons.

Some preliminary general considerations on the dynamics of the electronic
cloud are in order. First, we recall that, in previous work on damping
of a polyelectronic system in a Penning trap, the friction coefficient
of $Z$, $\gamma_{z}^{(N)}$, was shown to be well approximated as
$\gamma_{z}^{(N)}=N\gamma_{z}$ \cite{key-18,key-19}. Second, we
must take into account that the random force on the center-of-mass
coordinate is given by $\eta^{(N)}(t)=\frac{1}{N}\sum_{i=1}^{N}\eta_{i}(t)$,
where $\eta_{i}(t)$ denotes the noise on each individual electron.
The statistical characterization of $\eta^{(N)}(t)$ is straightforward.
As given by a linear superposition of Gaussian fluctuations, $\eta^{(N)}(t)$
has also Gaussian-noise characteristics. From the zero-mean values
of the mono-electronic stochastic forces, one trivially obtains $\left\langle \eta^{(N)}(t)\right\rangle =0$.
Additionally, assuming that the individual random forces are completely
uncorrelated, i.e., $\left\langle \eta_{i}(t)\eta_{j}(t^{\prime})\right\rangle =2D\delta_{ij}\delta(t-t^{\prime})$,
$i,j=1\ldots N$, we obtain $\left\langle \eta^{(N)}(t)\eta^{(N)}(t^{\prime})\right\rangle =2\frac{D}{N}\delta(t-t^{\prime})\equiv2D^{(N)}\delta(t-t^{\prime})$.
Hence,  the individual fluctuations average to a weaker noise in the
collective coordinate. (In order to present our arguments in simple
terms, we have considered here the white-noise limit. The generalization
to generic broadband noise is direct.) It follows that the main novelties
in the description of $Z$ with respect to the previously described
mono-electronic scenario are the presence of a larger damping coefficient
and of a reduced noise strength. Our analysis will focus on the implications
of those differential characteristics for the scaling of the phase-flip
rate with $N$. Since, as shown by the study of the one-electron oscillator,
the mechanism of stochastic phase switching does not essentially depend
on the anharmonicity of the potential, we can neglect the nonlinear
terms in the analysis. Accordingly, we consider the evolution of $Z$
as approximated by 

\begin{equation}
\ddot{Z}+\gamma_{z}^{(N)}\dot{Z}+\omega_{z}^{2}\left[1+h\cos\omega_{d}t\right]Z=\eta^{(N)}(t).\end{equation}

Given that Eq. (4) has the same structure as Eq. (1), the methodology
previously presented for the study of the mono-electronic system can
also be used here. Indeed, this parallelism allows us to extrapolate
some of the previous results. In particular, the flip rate, which,
for a one-electron system was found to scale as $\exp(-E/D)$, can
be expected now to have the form $\exp(-E^{(N)}/D^{(N)})$, where
$D^{(N)}$ is the (effective) strength of $\eta^{(N)}(t)$ and $E^{(N)}$
is the activation energy for $Z$. Crucial to the analysis is to take
into account that, both, $E^{(N)}$ and $D^{(N)}$, depend on $N$.
Actually, as pointed out in the above application of our approach,
the activation energy increases with the damping constant. Therefore,
as the friction coefficient is $\gamma_{z}^{(N)}=N\gamma_{z}$, we
conclude that $E^{(N)}$ increases with $N$, and, consequently, that
the phase flips are hindered as $N$ grows. An additional contribution
to the inhibition of the phase switching is rooted in the curbed fluctuations
in the center-of-mass coordinate: the decrease of the effective noise
strength $D^{(N)}=D/N$ with $N$ contributes also to the hindrance
of the phase jumps for growing electron number. We can go further
in the analytical characterization of the dependence of $\Gamma^{(N)}$
on $N$: by combining the equations $\Gamma^{(N)}\thicksim\exp(-E^{(N)}/D^{(N)})$,
$E^{(N)}=C_{1}+C_{2}N\gamma_{z}$, and $D^{(N)}=D/N$, we obtain \begin{equation}
\Gamma^{(N)}\thicksim\exp(-N(C_{1}+C_{2}N\gamma_{z})/D).\end{equation}
This expression is the key element in our explanation of the observed
scaling of the jump rate. The experimental procedure reported in Ref.
{[}4{]} included different variations of the system parameters. Specially
revealing of the mechanism responsible for the jump inhibition is
the analysis of the experimental run corresponding to a simultaneous
variation of $N$ and $\gamma_{z}$ with constant $N\gamma_{z}$.
Indeed, the observed decrease of the jump rate with $N$ points to
a mechanism not linked to the friction term $\gamma_{z}^{(N)}$, which,
in fact, is kept constant in this run. From our approach, we can conjecture
that, in this case, it is the decrease of the effective noise strength
that leads to the detected slow-down of the phase flips. More specifically,
the measured linear dependence of the exponent of $\Gamma^{(N)}$
on $N$ can be traced to the term $NC_{1}/D$ in our analytical characterization
of $\Gamma^{(N)}$ given by Eq. (5). Additional insight is provided
by the experimental results corresponding to a mere variation of $N$,
with constant $\gamma_{z}$. In this case, both, a reduction in $D^{(N)}$
and an increase of $\gamma_{z}^{(N)}$  take place. Since, again,
an approximately linear dependence of the exponent of $\Gamma^{(N)}$
on $N$ was found, we can conjecture that, in the regime studied in
the experiments, the reduced noise strength is the dominant element
in the mechanism responsible for the inhibition of the phase switching.
Following the report of the experimental findings, our discussion
in this section has focused on the persistence of the stochastic flips
in the polyelectronic system. For a more complete description of the
dynamics of the electronic cloud, the access to additional experimental
data is necessary.

\section{concluding remarks}

Our description of the stochastic dynamics of the one-electron Penning-trap
oscillator explains the experimental findings of Ref. {[}4{]}. The
physical mechanism responsible for the observed random phase switching
has been traced to the combination of a driving field at parametric
resonance and broadband noise entering additively the axial-mode equation.
The driving field allows the strong localization of the phase around
two equilibrium positions, and, therefore, the abrupt character of
the changes in phase. Additionally, the fluctuations establish a link
between the amplitude and the phase which results in a significant
enhancement of the effects of noise on the phase for small values
of the amplitude. Our analysis uncovers the generality of this mechanism,
and, therefore, its relevance to different contexts. Indeed, we have
reported its previous characterization in studies on the appearance
of selfexcited nonlinear oscillations in electronic devices \cite{key-17}.
Our work proves that the detected characteristics of the system response
do not specifically depend on the presence of residual colored noise
in the practical setup. In fact, it has been shown that the emergence
of the observed features can be simply traced to broadband fluctuations.
Finally, from a generalization of our approach, we have shown that
the observed attenuation of the phase flips in the dynamics of a cloud
of $N$ electrons can be explained as rooted in the effective reduction
of the noise strength in the center-of-mass coordinate.


\begin{thebibliography}{22}
\bibitem[1]{key-1}For a review, see L. S. Brown and G. Gabrielse,
Rev. Mod. Phys. \textbf{58}, 233 (1986), and references therein.

\bibitem[2]{key-2}G. Gabrielse \emph{et al.}, Phys. Rev. Lett. \textbf{100},
113001 (2008). 

\bibitem[3]{key-3}D. Hanneke \emph{et al.}, Phys. Rev. Lett. \textbf{100},
120801 (2008). 

\bibitem[4]{key-4} L. J. Lapidus \emph{et al.}, Phys. Rev. Lett.
\textbf{83}, 899 (1999). 

\bibitem[5]{key-5}S. Peil and G. Gabrielse, Phys. Rev. Lett. \textbf{83},
1287 (1999). 

\bibitem[6]{key-6}D. Enzer and G. Gabrielse, Phys. Rev. Lett. \textbf{78},
1211 (1997). 

\bibitem[7]{key-7}S. Brouard and J. Plata, Phys. Rev. A \textbf{64},
063405 (2001).

\bibitem[8]{key-8}B. Vestergaard and J. Javanainen, Phys. Rev. A
\textbf{58}, 1537 (1998). 

\bibitem[9]{key-9}J. Goldman, and G. Gabrielse, Phys. Rev. A \textbf{81},
052335 (2010). 

\bibitem[10]{key-10}J. L. Lamata \emph{et al.}, Phys. Rev. A \textbf{81},
022301 (2010). 

\bibitem[11]{key-11}C. H. Tseng \emph{et al.}, Phys. Rev. A \textbf{59},
2094 (1999).

\bibitem[12]{key-12}G. Ciaramicoli \emph{et al.}, Phys. Rev. A \textbf{70},
032301 (2004).

\bibitem[13]{key-13}M. I. Dykman \emph{et al.}, Phys. Rev. E \textbf{57},
5202 (1998).

\bibitem[14]{key-14}C. Stambaugh and H. B. Chan, Phys. Rev. B \textbf{73},
172302 (2006). 

\bibitem[15]{key-15}H. B. Chan and C. Stambaugh, Phys. Rev. Lett.
\textbf{99}, 060601 (2007). 

\bibitem[16]{key-16}N. N. Bogoliubov and Y. A. Mitropolsky, \emph{Asymptotic
Methods in the Theory of Non-Linear Oscillations} (Gordon and Breach,
New York, 1961).

\bibitem[17]{key-17} R. L. Stratonovich, \emph{Topics in the Theory
of Random Noise} (Gordon and Breach, New York, 1963).

\bibitem[18]{key-18}J. Tan and G. Gabrielse, Phys. Rev. Lett. \textbf{67},
3090 (1991).

\bibitem[19]{key-19}J. Tan and G. Gabrielse, Phys. Rev. A \textbf{48},
3105 (1993). 

\bibitem[20]{key-20}S. Brouard and J. Plata, Phys. Rev. A \textbf{65},
053412 (2002).

\bibitem[21]{key-21} H. Risken, \emph{The Fokker-Planck Equation}
(Springer-Verlag, New York, 1989)

\bibitem[22]{key-22}J. Plata, Phys. Rev. E \textbf{59}, 2439 (1999).
\end{thebibliography}
\end{document}